\documentclass[final,10pt,conference,a4paper]{IEEEtran}
\IEEEoverridecommandlockouts %

\usepackage{relsize}

\usepackage{array}
\usepackage{booktabs,tabularx}
\usepackage{multirow}
\usepackage{units}
\usepackage{graphicx}

\usepackage[T1]{fontenc}
\usepackage{textcomp}
\usepackage[utf8]{inputenc}
\usepackage[final]{microtype}
\usepackage{icomma}
\usepackage{xspace}

\usepackage[tbtags]{amsmath}
\usepackage{amssymb,amsfonts,bm}
\usepackage{mathtools} 
\usepackage{dsfont}

\usepackage{color}
\usepackage{tikz}
\usepackage{pgfplots}

\usepackage{algorithm}

\usepackage{algpseudocode}

\usepackage{glossaries}
\usepackage{ifthen}

\makeglossaries

\newacronym{mac}{MAC}{multiple-access channel}
\newacronym{bc}{BC}{broadcast channel}
\newacronym{mimo}{MIMO}{multiple-input multiple-output}
\newacronym{siso}{SISO}{single-input single-output}
\newacronym{af}{AF}{amplify-and-forward}
\newacronym{df}{DF}{decode-and-forward}
\newacronym{cf}{CF}{compress-and-forward}
\newacronym{mwrc}{MWRC}{multi-way relay channel}
\newacronym{pde}{PDE}{partial data exchange}
\newacronym{fde}{FDE}{full data exchange}
\newacronym{iid}{i.i.d.\@}{independent and identically distributed}
\newacronym{awgn}{AWGN}{additive white Gaussian noise}
\newacronym{awg}{AWG}{additive white Gaussian}
\newacronym{sic}{SIC}{successive interference cancellation}
\newacronym{snr}{SNR}{signal-to-noise ratio}
\newacronym{sinr}{SINR}{signal to interference plus noise ratio}
\newacronym{zf}{ZF}{zero-forcing}
\newacronym{mmse}{MMSE}{minimum mean square error}
\newacronym{sud}{SUD}{single user decoding}
\newacronym{dof}{DoF}{degrees of freedom}
\newacronym{gdof}{GDoF}{generalized degrees of freedom}
\newacronym{nnc}{NNC}{noisy network coding}
\newacronym{dmn}{DMN}{discrete memoryless network}
\newacronym{csi}{CSI}{channel state information}
\newacronym{ee}{EE}{energy efficiency}
\newacronym{ian}{IAN}{treating interference as noise}
\newacronym{snd}{SND}{simultaneous non-unique decoding}
\newacronym{pa}{PA}{power amplifier}
\newacronym{adc}{ADC}{analog-to-digital converter}
\newacronym{dac}{DAC}{digital-to-analog converter}
\newacronym{dsp}{DSP}{digital signal processing}

\makeglossaries

\newcommand{\eeq}{\end{equation}}

\newcommand{\bx}{\mbox{\boldmath $x$}}

\newcommand{\beq}{\begin{equation}}

\newcommand{\db}{{\mathbf d}}

\renewcommand{\bx}{\ensuremath{\bm{x}}\xspace}

\usetikzlibrary{positioning}
\usetikzlibrary{calc}
\usetikzlibrary{shapes.geometric}
\usetikzlibrary{shapes.multipart}
\usetikzlibrary{intersections}

\makeatletter
\newcommand\transformxdimension[1]{
    \pgfmathparse{((#1/\pgfplots@x@veclength)+\pgfplots@data@scale@trafo@SHIFT@x)/10^\pgfplots@data@scale@trafo@EXPONENT@x}
}
\newcommand\transformydimension[1]{
    \pgfmathparse{((#1/\pgfplots@y@veclength)+\pgfplots@data@scale@trafo@SHIFT@y)/10^\pgfplots@data@scale@trafo@EXPONENT@y}
}
\makeatother

\DeclareMathOperator\expval{\mathds{E}}
\newcommand{\abs}[1]{\ensuremath{\left\lvert #1 \right\rvert}}
\DeclareMathOperator\Capa{C}

\newcommand\SNR{\mathrm{SNR}}

\DeclareFontFamily{U}{mathx}{\hyphenchar\font45}
\DeclareFontShape{U}{mathx}{m}{n}{
      <5> <6> <7> <8> <9> <10>
      <10.95> <12> <14.4> <17.28> <20.74> <24.88>
      mathx10
      }{}
\DeclareSymbolFont{mathx}{U}{mathx}{m}{n}
\DeclareMathSymbol{\bigtimes}{1}{mathx}{"91}

\newtheorem{lemma}{Lemma}

\newtheorem{proposition}{Proposition}

\newtheorem{remark}{Remark}

\pgfplotscreateplotcyclelist{color2}{%
    blue,mark=*\\%
    red,mark=square*\\%
    brown!60!black,mark=triangle*\\%
    green!50!black,mark=diamond*\\%
    black,mark=star\\%
 }

\IEEEpubid{\makebox[\columnwidth]{978-1-4799-5863-4/14/\$31.00~\copyright~2014 IEEE \hfill} \hspace{\columnsep}\makebox[\columnwidth]{ }}

\begin{document}
\title{Spectral and Energy Efficiency in 3-Way Relay Channels with Circular Message Exchanges}

\author{\IEEEauthorblockN{Bho Matthiesen, Alessio Zappone, and Eduard A. Jorswieck}
\IEEEauthorblockA{Communications Theory, Communications Laboratory\\
Department of Electrical Engineering and Information Technology\\
Technische Universität Dresden, Germany\\
Email: \{bho.matthiesen, alessio.zappone, eduard.jorswieck\}@tu-dresden.de}%
\thanks{
The work of Bho Matthiesen and Eduard Jorswieck is supported by the German Research Foundation (DFG) in the
Collaborative Research Center 912 ``Highly Adaptive Energy-Efficient Computing.''

The work of Alessio Zappone has been funded by the German Research Foundation (DFG) project CEMRIN, under grant ZA 747/1-2.}}

\maketitle

\begin{abstract}
Spectral and energy efficiency in 3-way relay channels are studied in this paper. First, achievable sum rate expressions for 3-way relay channels are derived for different relaying protocols. Moreover, an outer bound for the capacity of the 3-way relay channel is presented. Next, leveraging the derived achievable sum rate expressions, two algorithms for joint power allocation at the users and at the relay are designed so as to maximize the system energy efficiency. Numerical results are provided to corroborate and provide insight on the theoretical findings. 
\end{abstract}


\IEEEpeerreviewmaketitle

\section{Introduction}
Relays are fundamental building blocks of wireless networks. One recently proposed channel model for relay networks is the \gls{mwrc}. Such a model applies to many communication architectures like the communication of several ground stations over a satellite, or wireless board-to-board communication in highly adaptive computing \cite{FettDATE13} where multiple chips exchange data with the help of another chip acting as relay. The \gls{mwrc} was first introduced in \cite{Gunduz2013}, where all users in the cluster send a message and are interested in decoding the messages of all other users in the cluster.
In \cite{Ong2010} the common-rate capacity of the \gls{awgn} \gls{mwrc} with full message exchange is given and it is shown that for three and more users this capacity is achieved by \gls{df} for \glspl{snr} below \unit[0]{dB} and compute-and-forward otherwise. In \cite{Chaaban2013} a constant gap approximation of the capacity region of the Gaussian 3-user \gls{mwrc} with full message exchange is given.

Besides spectral efficiency, another key performance metric in modern and future 5G wireless networks is \gls{ee}. From a mathematical standpoint, one well-established definition of the \gls{ee} of a communication system is the ratio between the system capacity or achievable rate and the total consumed power \cite{Miao2011,Isheden2012}. With this definition, the \gls{ee} is measured in bit/Joule. Previous results on \gls{ee} in relay systems mainly focus on regular \gls{af} or \gls{df} schemes and do not consider the \gls{mwrc}. In \cite{Zhou2011} the optimal placement of relays in cellular networks is investigated and is seen to provide power-saving gains. \cite{ZapTWC13} considers the bit/Joule definition of \gls{ee} and devises energy-efficient power control algorithms in interference networks. A cooperative approach is considered in \cite{ZapTSP13}, where a \gls{mimo} \gls{af} relay-assisted system is considered.

In this paper a 3-way relay channel is considered and both spectral and energy efficiency are analyzed and optimized.
In contrast to most other works on \glspl{mwrc}, we focus on a partial message exchange where each message is only destined for one receiver and, also, not every user sends a message to each other user. This makes it necessary to deal with interference at the receivers which complicates the analysis. However, it might also result in higher achievable rates due to less decoding constraints.
The contributions of the paper can be summarized as follows: 1) achievable sum rate expressions are derived for the \gls{af}, \gls{df}, and \gls{nnc} relaying protocols; 2) an outer bound for the capacity of the 3-way relay channel is derived and used for benchmarking purposes; 3) building on the derived achievable sum rate expressions, two algorithms for energy efficiency optimization are provided to jointly allocate the users' and the relays transmit powers.

We define the function $C(x) = \log_2(1+x)$ for $x\ge 0$.

\section{System Model}
We consider the symmetric 3-user \gls{siso} \gls{mwrc} with circular (i.e.,
partial) message exchange, \gls{awgn}, no direct user-to-user links, and full-duplex
transmission. The users are denoted as node 1 to 3 and the relay is node 0. We
define the set of all users as $\mathcal K = \{ 1, 2, 3 \}$ and the set of all
nodes as $\mathcal K_0 = \mathcal K \cup \{ 0 \}$.

The 3-user \gls{mwrc} consists of an uplink channel
	$Y_0 = \sum_{k\in\mathcal K} X_i + Z_0$,
and downlink channels
	$Y_k = X_0 + Z_k,\quad k\in\mathcal K$,
where $X_k$ and $Y_k$ are the complex valued channel input and output at node
$k\in\mathcal K_0$, respectively, and $Z_k$ is zero mean circularly symmetric
complex Gaussian noise with power $N_0$ at the relay and $N$ at all other
nodes.

All noise variables are mutually independent and the channel inputs are
\acrlong{iid} over time. All channel inputs have zero mean and an average power
constraint $\expval \abs{X_0}^2 \le P_0$
and $\expval \abs{X_k}^2 \le P$, for $k\in\mathcal K$.

\begin{figure}
	\begin{center}
		\begin{tikzpicture}
			\begin{scope}[every node/.style=draw,circle]
				\node (n1) {1};
				\node (n2) at (0:3.5) {2};
				\node (n3) at (60:3.5) {3};
				\node (nR) at ($1/3*(n1)+1/3*(n2)+1/3*(n3)$) {0};
			\end{scope}

			\foreach \a/\da/\db in {1/dotted/dashed,2/dashdotdotted/dotted,3/dashed/dashdotdotted} { 
				\begin{scope}[thick]
					\draw [\da,latex-,relative,out=350,in=190] (nR) to (n\a);
					\draw [\db,-latex,relative,out=10,in=170] (nR) to (n\a);
				\end{scope}
			}
		\end{tikzpicture}
	\end{center}
	\vspace{-1em}
	\caption{Illustration of the system model where node 0 is the relay and nodes 1 to 3 are the users. Messages travel along the different line styles.}
	\label{fig:msgex12}
\end{figure}
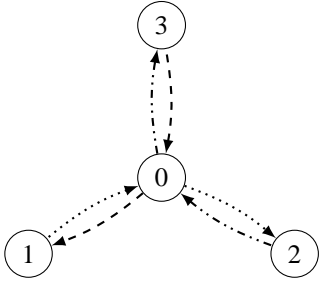

We consider a circular message exchange as illustrated in Fig.~\ref{fig:msgex12} where
user $q(k)$ wants message $m_{k}$ with $q = [ 2, 3, 1 ]$. We also define $l(k)$ as the index of the interfering (i.e., unwanted) message at user $q(k)$ as $l = [3, 1, 2]$.

A $(2^{nR_1}, 2^{nR_2}, 2^{nR_3}, n)$ code for the 3-user \gls{mwrc}
consists of
		three message sets $\mathcal M_k = [1:2^{nR_k}]$, one for each user
		$k\in\mathcal K$,
		three encoders, where encoder $k\in\mathcal K$ assigns a symbol
		$x_{ki}(m_k, y_k^{i-1})$ to each message $m_k \in\mathcal M_k$ and
		received sequence $y_k^{i-1}$ for $i\in[1:n]$,
		a relay encoder that assigns a symbol $x_{0i}(y_0^{i-1})$ to every
		past received sequence $y_0^{i-1}$ for $i\in[1:n]$, and
		three decoders, where decoder $k\in\mathcal K$ assigns an
		estimate $\hat m_{q(k)} \in \mathcal M_{q(k)}$ or an error message $e$
		to each pair $(m_k, y_k^n)$.

We assume that the message triple $(M_1, M_2, M_3)$ is uniformly distributed over
$\mathcal M_1 \times \mathcal M_2 \times \mathcal M_3$. The average probability
of error is defined as
	$P_e^{(n)} = \Pr\left\{ \hat M_k \neq M_k \text{ for some } k\in\mathcal K \right\}$.

A rate triple $(R_1, R_2, R_3)$ is said to be achievable if there exists a
sequence of $(2^{nR_1}, 2^{nR_2}, 2^{nR_3}, n)$ codes such that
$\lim_{n\rightarrow\infty} P_e^{(n)} = 0$. The capacity region of the 3-user
\gls{mwrc} is the closure of the set of achievable rates. The sum rate is
defined as $R_\Sigma = \max\left\{ R_1 + R_2 + R_3 : \left( R_1, R_2, R_3
\right) \in \mathcal R \right\}$, where $\mathcal R$ is an achievable rate
region. Whenever $\mathcal R$ is the capacity region, we call $R_\Sigma$ the
sum capacity $C_\Sigma$.

\section{Bounds on the Sum Capacity}\label{Sec:AchievableRates}
We start our treatment of the symmetric 3-user \gls{mwrc} by deriving an upper
bound on the sum capacity and then continue with several inner bounds.

\subsection{Outer Bound}
This outer bound consists of the cut set bound in the
uplink and a downlink bound \cite{Yoo2009} that takes the side information
at the receivers into account.
\begin{lemma} \label{lem:outerbound}
	The sum capacity of the symmetric 3-user \gls{mwrc} is upper bounded as
	\begin{equation} \label{eq:outerbound}
		C_\Sigma \le \min\left\{ \frac{3}{2} \Capa\left( \frac{P_0}{N} \right),\ 3 \Capa\left( \frac{P}{N_0} \right) \right\}.
	\end{equation}
\end{lemma}
\begin{IEEEproof}
	The proof is omitted due to space constraints.
\end{IEEEproof}

\subsection{Amplify-and-Forward}
We first consider \gls{af} relaying where the relay scales the observed signal by
a positive constant and broadcasts it back to the users. The transmitted
symbol at the relay is
	$X_0 = \alpha Y_0$,
where $\alpha$ is a normalization factor chosen such that the transmit power
constraint at the relay is met, i.e.,
$\alpha = \sqrt{P'_0\,/ \left( \sum_{k\in\mathcal K} P'_k + N_0 \right)}$,
where $P'_k$, $k\in\mathcal K_0$, is the actual transmit power of node $k$
satisfying the average power constraints.
The receiver first removes its self-interference from the received signal and then decodes for its desired message while treating the remaining interference as noise.

We split the transmission into three equal length blocks and switch off user
$i$ in time slot $i$. This reduces interference and allows for higher transmission powers in the other
two time slots while still meeting the average power constraint.

\begin{lemma} \label{lem:af}
	In the 3-user \gls{mwrc}, the sum rate
	\begin{equation} \label{eq:af}
		R_\Sigma^{AF} = \Capa\left( \frac{3 P P_0}{N_0 P_0 + 3 P N + N N_0} \right)
	\end{equation}
	is achievable with \gls{af} relaying and treating interference as noise at the receivers.
\end{lemma}
\begin{IEEEproof}
	Omitted due to space limitations.
\end{IEEEproof}
\subsection{Decode-and-Forward}
In \gls{df} relaying, the relay completely decodes the messages of each user
and then broadcasts them back to all users. The achievable rate region is the
intersection of the capacity regions of the 3-user \acrlong{mac} and the \acrlong{bc}
with receiver side information and partial decoding at the receivers.

\begin{lemma} \label{lem:df}
	In the 3-user \gls{mwrc}, the sum rate
	\begin{equation} \label{eq:dfsr}
		R_\Sigma^{DF} = \min\left\{ \frac{3}{2} \Capa\left( \frac{P_0}{N} \right),\ \Capa\left( \frac{3 P}{N_0} \right) \right\}
	\end{equation}
	is achievable with \gls{df} relaying.
\end{lemma}
\begin{IEEEproof}[Proof sketch]
	The achievable rate region is given in \cite[Proposition 2]{Gunduz2013}.
	Using the simplex algorithm and the fact that
	$\Capa\left( \frac{2 P}{N_0} \right) < \Capa\left( \frac{P_0}{N} \right)$
	implies
	$\Capa\left( \frac{3 P}{N_0} \right) < \frac{3}{2} \Capa\left( \frac{P_0}{N} \right)$
	we can prove \eqref{eq:dfsr}.
\end{IEEEproof}

\begin{remark}
	The result from \cite{Gunduz2013} implements a full message exchange.
	However, from the outer bound in \cite{Yoo2009} it can be seen that in the
	symmetric case the relaxed decoding requirements due to the partial message
	exchange considered here can not result in higher rates for \gls{df}.
\end{remark}

\begin{remark} \label{rem:caplowsnr}
	For a completely symmetric scenario, \gls{df} is sum rate optimal in the low \gls{snr} regime.
	To see this, let $P=P_0$ and
	$N=N_0$ and define $S = \frac{P}{N}$. Then the bound in Lemma~\ref{lem:outerbound} is
	$C_\Sigma \le \frac{3}{2} \Capa\left( S \right)$,
	and
	$R_\Sigma^{DF} = \min\left\{ \frac{3}{2} \Capa\left( S \right),\ \Capa\left( 3 S \right) \right\}$.
	It is easily shown that for $S \le 3 + 2 \sqrt{3}$ the first term in the minimum
	is dominant, i.e., $R_\Sigma^{DF} = \frac{3}{2}
	\Capa\left( S \right)$. Since this is equal to
	the outer bound, we have $C_\Sigma = \frac{3}{2} \Capa\left( S \right)$ for
	\glspl{snr} up to $3 + 2 \sqrt{3} \approx \unit[8.1]{dB}$.
\end{remark}

\subsection{Noisy Network Coding}
\Gls{nnc} \cite{Lim2011} generalizes \acrlong{cf} to \acrlongpl{dmn}. For general
multi-message networks there are two different decoding methods to choose from:
\acrfull{snd} and \acrfull{ian}. Since, in general, none of the methods is
superior to the other, we evaluate both bounds. However, it turns out that
\gls{ian} is strictly worse than \gls{snd} and even than \gls{af}.

\begin{lemma} \label{lem:nnc-snd}
	In the 3-user \gls{mwrc}, the sum rate
	\begin{equation} \label{eq:nnc-snd}
		R_\Sigma^{NNC-SND} = \frac{3}{2} \Capa\left( \frac{2 P P_0}{N_0 P_0 + 2 P N + N N_0} \right)
	\end{equation}
	is achievable with \gls{nnc} and simultaneous non-unique decoding.
\end{lemma}
\begin{IEEEproof}[Proof sketch]
	Use \cite[Theorem 2]{Lim2011} and identify $\mathcal D_0 = \emptyset$ and
	$\mathcal D_k = \{ q(k) \}$ for $k\in\mathcal K$. Assume $\hat Y_i = Y_i +
	\hat Z_i$ with $\hat Z_i \sim \mathcal{CN}(0, Q_i)$ for $i\in\mathcal K_0$,
	and $Q = \emptyset$, i.e., no time-sharing is used.
	Then, the achievable rate region is
	\begin{align*}
		R_k &< \Capa\left( \frac{P}{N_0 + Q_0} \right),\\
		\smashoperator{\sum_{i\in\mathcal K\setminus\{k\}}} R_i &< \min \left\{ \Capa\left( \frac{2 P}{N_0 + Q_0} \right),\ \Capa\left( \frac{P_0}{N} \right) - \Capa\left( \frac{N_0}{Q_0} \right) \right\},
	\end{align*}
	for each $k\in\mathcal K$.

	With the simplex algorithm \cite{Luenberger2008} and after maximization over $Q_0$ we get \eqref{eq:nnc-snd}.
\end{IEEEproof}

\begin{remark} \label{rem:AFworseSND}
	It can be shown that $R_\Sigma^{NNC-SND} \ge R_\Sigma^{AF}$.
\end{remark}
\begin{lemma} \label{lem:nnc-ian}
	In the 3-user \gls{mwrc}, the sum rate
	\begin{equation*}
		R_\Sigma^{NNC-IAN} = 3 \Capa\left( \frac{P P_0}{2 P P_0 + N_0 P_0 + 3 P N + N N_0} \right)
	\end{equation*}
	is achievable with \gls{nnc} and treating interference as noise.
\end{lemma}
\begin{IEEEproof}
	Similar to Lemma~\ref{lem:nnc-snd}.
\end{IEEEproof}
\begin{remark} \label{rem:IANworseAF}
	It can be shown that $\strut R_\Sigma^{NNC-IAN} \le R_\Sigma^{AF}$. Thus, with
	Remark~\ref{rem:AFworseSND}, $R_\Sigma^{NNC-IAN} \le R_\Sigma^{NNC-SND}$.
\end{remark}

\section{Energy Efficiency}
\label{sec:EE}
The \gls{ee} of the system is defined as the ratio between the achievable sum rate and the consumed power. The power consumed in the system is given by the sum of the transmit power of each user and of the relay plus the circuit power that is dissipated in each terminal to operate the devices. Moreover, the transmit power of each terminal should be scaled by a factor larger than $1$ to model the nonidealities of the power amplifier \cite{Isheden2012,circuit_power_consumption}. Namely, we can express the total power $P_{t}$ consumed in the network as $P_{t}=\phi P+\psi P_{0}+P_{c}$, with $P_{c}$ denoting the total circuit power consumed in all nodes, $\psi\geq 1$ being the inefficiency of the relay amplifier, and $\phi\geq 3$, accounting for the inefficiency of the amplifier of the three users.
Accordingly, the \gls{ee} can be defined as the ratio between the achievable sum rate and $P_{t}$.
Then, given the achievable sum rate expressions from Section~\ref{Sec:AchievableRates}, the \gls{ee} can be expressed in two different functional forms.
For the upper-bound and for the \gls{df} case we have 
\[
{\rm EE}_{1}=\displaystyle\frac{\min\left\{a_{1}\Capa\left(\alpha_{1}\frac{P_{0}}{N}\right),\ a_{2}\Capa\left(\alpha_{2}\frac{P}{N_{0}}\right)\right\}}{\phi P+\psi P_{0}+P_{c}}\;,
\]
with $a_{1}$, $a_{2}$, $\alpha_{1}$, and $\alpha_{2}$ non-negative parameters.
For the \gls{af} and \gls{nnc} cases we have
\[
{\rm EE}_{2}=\displaystyle\frac{\alpha\Capa\left(\frac{PP_{0}}{aP+bP_{0}+c}\right)}{\phi P+\psi P_{0}+P_{c}}\;,
\]
with $\alpha$, $a$, $b$, and $c$ non-negative parameters.
In the following, \gls{ee} maximization will be carried out by means of fractional programming tools. In particular, we recall the following result from \cite{FracProgSS1983,NonlinearFracProg}. Consider the generic fractional problem 
$%
\mathop{\max}_{\bx\in\mathcal{S}} \frac{f(\bx)}{g(\bx)}
$ %
where ${\cal S}\in\mathbb{R}^n$, $f,g:\mathcal{S}\to\mathbb{R}$, with $f(\bx)\geq 0$ and $g(\bx)>0$. Define the function
$F(\lambda)=\mathop{\max}\limits_{\bx\in\mathcal{S}}\left(f(\bx)-\lambda g(\bx)\right)$.
Then, maximizing $f(\bx)/g(\bx)$ is equivalent to finding the unique zero of $F(\lambda)$. This can be accomplished by means of Dinkelbach's algorithm \cite{NonlinearFracProg}, which only requires the solution of a sequence of convex problems, provided $f(\bx)$ and $g(\bx)$ are concave and convex, respectively, and that ${\cal S}$ is a convex set.
Moreover, it can be shown that the convergence rate of Dinkelbach's algorithm is superlinear \cite{NonlinearFracProg}. 

\subsection{Maximization of ${\rm EE}_{1}$}
The maximization of ${\rm EE}_{1}$ is a non-concave and non-smooth problem. However, it can be reformulated as a smooth problem introducing the auxiliary variable $t$ as follows.
\beq\label{Prob:MaxEE1Smooth}
\left\{
\begin{array}{lll}
\displaystyle\max_{P,P_{0}}\quad \frac{t}{\phi P+\psi P_{0}+P_{c}}\\
{\rm s.t.}\;\;\; P\in[0;P^{max}]\;,\;P_{0}\in[0;P_{0}^{max}]\\
a_{1}\Capa\left(\alpha_{1}\frac{P_{0}}{N}\right)-t\geq 0\;,\quad a_{2}\Capa\left(\alpha_{2}\frac{P}{N_{0}}\right)-t \geq 0
\end{array}
\right.
\eeq
The numerator and denominator of the objective of \eqref{Prob:MaxEE1Smooth} are both linear and the constraints are convex. As a consequence, \eqref{Prob:MaxEE1Smooth} can be solved by means of Dinkelbach's algorithm with an affordable complexity.

\subsection{Maximization of ${\rm EE}_{2}$}
In this case, the optimization problem is formulated as
\beq\label{Prob:MaxEE2}
\left\{
\begin{array}{ll}
\displaystyle\max_{P,P_{0}}\displaystyle\frac{\alpha\Capa\left(\frac{PP_{0}}{aP+bP_{0}+c}\right)}{\phi P+\psi P_{0}+P_{c}}\\
{\rm s.t.}\;\;\; P\in[0;P^{max}]\;,\quad P_{0}\in[0;P_{0}^{max}]
\end{array}
\right.
\eeq
Problem \eqref{Prob:MaxEE2} is more challenging than Problem \eqref{Prob:MaxEE1Smooth} because the numerator of the objective function is not jointly concave in the optimization variables.
However, we observe that the numerator of the objective is separately concave in $P$ for fixed $P_{0}$ and vice versa. This suggests that a convenient way to tackle Problem \eqref{Prob:MaxEE2} is by means of the alternating maximization algorithm \cite{BertsekasNonLinear}, according to which we can alternatively optimize with respect to $P$ fixing the value of $P_{0}$, and with respect to $P_{0}$ for a fixed value of $P$.

The formal algorithm is reported next and labeled Algorithm 1.
\begin{algorithm}
\caption{Alternating maximization for Problem \eqref{Prob:MaxEE2}}
\label{Alg:AMEE2}
\begin{algorithmic}
\State
\texttt{Initialize} $P_{0}^{(0)}\in[0,P_{0}^{max}]$. \texttt{Set a tolerance} $\epsilon$. \texttt{Set} $n=0$;
\While{$\left|{\rm EE}_{2}^{(n)}-{\rm EE}_{2}^{(n-1)}\right| \leq \epsilon$}
\State \texttt{Given} $P_{0}^{(0)}$, \texttt{solve Problem} \eqref{Prob:MaxEE2} \texttt{with respect to} $P$ \texttt{to obtain the optimal} $P^{(n+1)}$;
\State \texttt{Given} $P^{(n+1)}$, \texttt{solve Problem} \eqref{Prob:MaxEE2} \texttt{with respect to} $P_{0}$ \texttt{to obtain the optimal} $P_{0}^{(n+1)}$;
\State $n=n+1$;
\EndWhile
\State \texttt{Output} $(P,P_{0})$.
\end{algorithmic}
\end{algorithm} 

Each subproblem in Algorithm~\ref{Alg:AMEE2} can be globally solved by means of Dinkelbach's algorithm.
Moreover, the following proposition holds.
\begin{proposition}
Algorithm~\ref{Alg:AMEE2} converges to a stationary point of Problem \eqref{Prob:MaxEE2}.
\end{proposition}
\begin{IEEEproof}[Proofsketch]
After each iteration of Algorithm~\ref{Alg:AMEE2} the objective is not decreased. Hence, convergence follows since the objective is upper-bounded. 
Convergence to a stationary point holds by virtue of \cite[Proposition 2.7.1]{BertsekasNonLinear}, which states that alternating maximization converges to a stationary point if: 1) the feasible set is the Cartesian product of closed and convex sets; 2) the objective is continuously differentiable on the feasible set; 3) the solution to each subproblem is unique. In our case, 1) and 2) are apparent. As for 3) it also holds because the objective function of each subproblem can be shown to be strictly pseudo-concave \cite{ZapTWC13}.
\end{IEEEproof}

\section{Discussion \& Numerical Results}
For a discussion and numerical evaluation of the presented transmission
schemes, we consider a completely symmetric scenario with $N = N_0$. We assume
$P = P_0$ for the spectral efficiency, and for the \gls{ee} evaluation, we
assume $P^{max} = P_0^{max}$, unit noise variance and no power loss at the
transmitter, i.e., $\psi = 1$ and $\phi = 3$. The shown performance has been
obtained using the algorithms proposed in Section~\ref{sec:EE}.

Fig.~\ref{fig:spectral_eff} shows the achievable sum rates from
Section~\ref{Sec:AchievableRates} as a function of the \gls{snr}. As noted in
Remarks~\ref{rem:AFworseSND} and \ref{rem:IANworseAF}, it can be observed that \gls{nnc} with \gls{snd}
achieves a higher sum rate than \gls{af} and \gls{af} achieves a higher sum
rate than \gls{nnc} with \gls{ian}. In the low \gls{snr} regime, the sum
capacity is achieved by \gls{df} (see Remark~\ref{rem:caplowsnr}). Starting at
approximately \unit[8]{dB}, \gls{df} stops being sum rate optimal but is still
better than all other considered transmission schemes. Starting from approximately
\unit[14.27]{dB} \gls{nnc} \gls{snd} is the best in terms of spectral efficiency.
Its gap to the outer bound is at most $\unit[1.5 \log_2\left( 1.5 \right)]{bit}
\approx \unit[0.877]{bit}$. In contrast, for all other considered schemes this
gap grows unbounded as $\SNR\rightarrow\infty$. Furthermore, the gap between
\gls{df} and \gls{af} approaches \unit[2]{bit} as $\SNR\rightarrow\infty$.
\Gls{nnc} \gls{ian} is clearly worse than every other employed scheme and
should not be considered in terms of spectral efficiency.

Fig.~\ref{fig:ee} shows the \gls{ee} as a function of the \gls{snr} for a
fixed circuit power $P_c = \unit[1]{W}$. First of all, it can be seen that the \gls{ee} saturates when $P_{max}$ exceeds a given value, which is lower than \unit[0]{dB} for all considered schemes. This is explained recalling that, unlike the achievable rate, the \gls{ee} is not increasing with the transmit powers, but instead admits an optimum transmit power level. If $P_{max}$ is larger than such power level, then it is not optimal to transmit at full power. This also explains why \gls{df} performs significantly better than all other schemes, including \gls{nnc} \gls{snd}. Indeed, due to the saturation of the \gls{ee}, the \gls{snr} range for which \gls{nnc} \gls{snd} yields a larger achievable sum rate than \gls{df} is not reached when \gls{ee} is optimized. Finally, as expected, \gls{nnc} \gls{snd} is better than \gls{af}, which is better than \gls{nnc} \gls{ian}.

However, as opposed to \gls{df}, \gls{af} does not require power-hungry analog
to digital conversion and digital signal processing at the relay which results
in significantly less power consumption. Furthermore, the decoders at the users
are also expected to consume less power due to the use of a (considerably
simpler) single user receiver.  Thus, the higher achievable rates and the
resulting better \gls{ee} of \gls{df} over \gls{af} are obtained at the cost of a more complex
hardware and, hence, of a larger consumed circuit power. This suggests that the
comparison in Fig.~\ref{fig:ee} might be unfair and that the large gap between
\gls{df} and \gls{af} might in fact be smaller when the comparison is done on
equal grounds. Some insight on this issue is given in Fig.~\ref{fig:eeAFDF}, which shows the
\gls{ee} of \gls{df} as a function of its circuit power $P_c^{DF}$ and the
\gls{ee} of \gls{af} for a fixed circuit power $P_c^{AF} = \unit[1]{W}$. It can
be seen that, as expected, the gap to \gls{af} gets smaller with increasing
$P_c^{DF}$ and that \gls{af} might even outperform \gls{df} given a
significantly large $P_c^{DF}$.

\tikzset{/pgfplots/width = {\axisdefaultwidth}, /pgfplots/height= {0.85*\axisdefaultheight}}

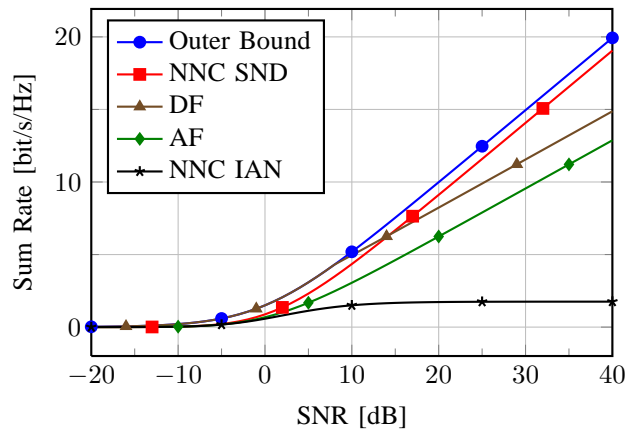
\begin{figure}
\begin{tikzpicture}
	\begin{axis} [
			thick,
			xlabel={SNR [dB]},
			ylabel={Sum Rate [bit/s/Hz]},
			ylabel near ticks,
			grid=major,
			xtick = {-20,-10,...,40},
			xmin = -20,
			xmax = 40,
			minor x tick num = 1,
			minor y tick num = 1,
			yminorgrids = true,
			mark repeat = 15,
			legend pos=north west,
			legend cell align=left,
			cycle list name=color2,
		]

		\pgfplotstableread{sym_sumrates.dat}\tbl

		\addplot+[mark phase = 0] table[y=bound] {\tbl};
		\addlegendentry{Outer Bound};

		\addplot+[mark phase = 8] table[y=nnc_snd] {\tbl};
		\addlegendentry{NNC SND};

		\addplot+[mark phase = 5] table[y=df] {\tbl};
		\addlegendentry{DF};

		\addplot+[mark phase = 11] table[y=af] {\tbl};
		\addlegendentry{AF};

		\addplot+[mark phase = 0] table[y=nnc_ifn] {\tbl};
		\addlegendentry{NNC IAN};
	\end{axis}
\end{tikzpicture}
\vspace{-2ex}
\caption{Spectral efficiency in the 3-user \gls{mwrc}; 1) Outer bound from
Lemma~\ref{lem:outerbound}, 2) \acrfull{nnc} with
\acrfull{snd} and 3) \acrfull{ian}, 4) \acrfull{af} and 5) \acrfull{df} plotted as a
function of the \gls{snr}.}
\label{fig:spectral_eff}
\end{figure}

\begin{figure}
\begin{tikzpicture}
	\begin{axis} [
			thick,
			xlabel={$P^{max}$ [dB]},
			ylabel={Energy Efficiency [bit/Hz/J]},
			ylabel near ticks,
			grid=major,
			xtick = {-40,-30,...,20},
			xmin = -30,
			xmax = 10,
			minor x tick num = 1,
			minor y tick num = 1,
			yminorgrids = true,
			mark repeat = 150,
			legend pos=north west,
			legend cell align=left,
			cycle list name=color2,
		]

		\pgfplotstableread{out_Pc=1_N=1.dat}\tbl

		\addplot+[mark phase = 0] table[y=Outer_Bound] {\tbl};
		\addlegendentry{Outer Bound};

		\addplot+[mark phase = 15] table[y=NNC_SND] {\tbl};
		\addlegendentry{NNC SND};

		\addplot+[mark phase = 90] table[y=DF] {\tbl};
		\addlegendentry{DF};

		\addplot+[mark phase = 105] table[y=AF] {\tbl};
		\addlegendentry{AF};

		\addplot+[mark phase = 60] table[y=NNC_IFN] {\tbl};
		\addlegendentry{NNC IAN};
	\end{axis}
\end{tikzpicture}
\vspace{-2ex}
\caption{\Acrlong{ee} in the 3-user \gls{mwrc} of 1) \acrfull{nnc} with
\acrfull{snd} and 2) \acrfull{ian}, 3) \acrfull{af}, 4) \acrfull{df}, and 5) the outer
bound from Lemma~\ref{lem:outerbound} as a function of the \gls{snr} for fixed
circuit power $P_c = \unit[1]{W}$.}
\label{fig:ee}
\end{figure}
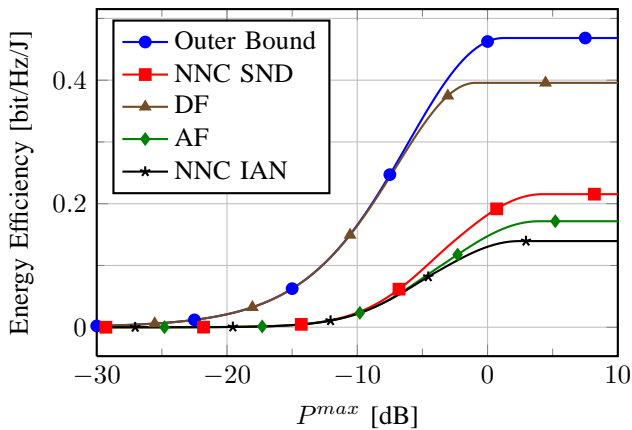

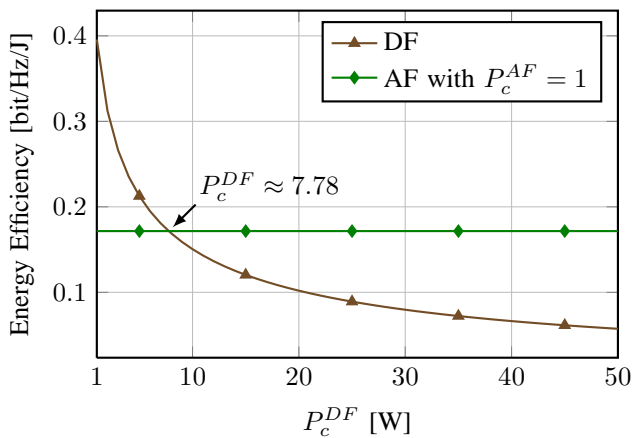
\begin{figure}
\begin{tikzpicture}
	\begin{axis} [
			thick,
			xlabel={$P_c^{DF}$ [W]},
			ylabel={Energy Efficiency [bit/Hz/J]},
			ylabel near ticks,
			grid=major,
			xtick = {1, 10, 20, 30, 40, 50},
			xmin = 1,
			xmax = 50,
			minor x tick num = 1,
			mark repeat = 10,
			legend pos=north east,
			legend cell align=left,
			cycle list name=color2,
			cycle list shift=2,
		]

		\pgfplotstableread{DFstepPc.dat}\tbl
		
		\pgfplotstablegetelem{1}{AF}\of\tbl
		\let\AFval\pgfplotsretval

		\addplot +[name path global=steppcDF, mark phase=5] table[y=DF] {\tbl};
		\addlegendentry{DF};

		\addplot +[name path global=steppcAF, mark phase=5] table[y=AF] {\tbl};

		\addlegendentry{AF with $P_c^{AF} = 1$};

		\path [inner sep=.1em, name intersections={of=steppcDF and steppcAF}]
			(intersection-1) \pgfextra{\pgfgetlastxy{\macrox}{\macroy}\let\macrox\macrox}
			node (int) {}
		 	+(1em, 1em)
			node (text) [anchor=south west] {$P_c^{DF} \approx \transformxdimension{\macrox}\pgfmathprintnumber[fixed,precision=2]{\pgfmathresult}$}
			\pgfextra{\transformxdimension{\macrox}\global\let\Pcintersection\pgfmathresult};

		\draw[latex-] (int.north east) -- (text.south west);
	\end{axis}
\end{tikzpicture}
\vspace{-2ex}
\caption{\Acrlong{ee} in the 3-user \gls{mwrc} of \acrfull{df} as a function of
the circuit power $P_c^{DF}$ compared to \acrfull{af} with a fixed circuit power
$P_c^{AF} = \unit[1]{W}$ for an operating point of $P^{max} = \unit[10]{W}$. The intersection is at $P_c^{DF}
\approx \pgfmathprintnumber[fixed,precision=2]{\Pcintersection}\, W$.}
\label{fig:eeAFDF}
\end{figure}

\section{Conclusion}
In this paper, we studied both the spectral and the energy efficiency of the
3-user \gls{mwrc} with a partial message exchange. We provided analytic sum
rate expressions for the most common relaying schemes and discussed the solution of the
optimization problems arising in the calculation of the \gls{ee}.

\enlargethispage{-4.6cm}

We have seen that if we assume the same power consumption for all schemes,
\gls{df} performs best in terms of \gls{ee}.  Moreover, the
energy-efficient  performance of \gls{nnc} is not satisfactory due to the
fact that \gls{nnc} achieves a higher spectral efficiency only in the high
\gls{snr} regime, which is not the operating regime when \gls{ee} is
optimized.  Furthermore, we have shown that \gls{af} might have better \gls{ee}
than the more complex \gls{df} if different circuit powers are assumed. This
assumption is reasonable since different hardware complexities imply different
circuit powers. Thus, to compare the \gls{ee} of the presented relaying schemes
in a fairer way, circuit power consumption models are necessary. This issue
will be addressed further in future work.

\bibliography{IEEEabrv,mwrc_ee.bib}

\end{document}